# Development of various methods for PrF$_3$ nanoparticles synthesis

E.M. Alakshin[a], B.M. Gabidullin[b], A.T. Gubaidullin[b], A.V. Klochkov[a], S.L. Korableva[a], M.A. Neklyudova[c], A.M. Sabitova[a], M.S. Tagirov[a]

[a] *Kazan (Volga region) Federal University, Kremlevskaya, 18, 420008, Kazan, Russia*
*alakshin@gmail.com*
[b] *A.E. Arbuzov Institute of Organic and Physical Chemistry of Kazan Scientific Center of Russian Academy of Sciences, Ac. Arbuzov street, 420088, Kazan, Russia*
[c] *A.V. Rzhanov Institute of Semiconductor Physics, pr. Lavrentieva,13, 630090, Novosibirsk, Russia*

*submitted to arXiv April 1$^{st}$, 2011*

**Abstract** The six nanosized PrF$_3$ samples were synthesized using two different chemical reactions and different time of hydrothermal reaction. The X-ray and HRTEM experiments showed high crystallinity of synthesized samples. For all samples the particles size distribution was obtained. It was shown, that precursors of chemical reaction have influence on the shape of synthesized nanoparticles. The size of nanoparticles depended on the time of hydrothermal reaction as much as roughly 10 nm per hour.

**Keywords** PrF3, nanoparticles, synthesis, van Vleck

**Introduction**

The "PrF$_3$–liquid $^3$He" system is of interest because of the possibility of using the magnetic coupling between the nuclei of the two spin systems for the dynamic nuclear polarization of liquid $^3$He. Van Vleck paramagnets are known to have high anisotropy of the effective nuclear magnetogyric ratio and for $^{141}$Pr in PrF$_3$ $\gamma_x/2\pi$=3.322 kHz/Oe, $\gamma_y/2\pi$=3.242 kHz/Oe, $\gamma_z/2\pi$=10.03 kHz/Oe, while for $^3$He $\gamma/2\pi$ = 3.243 kHz/Oe. As a result, a direct interaction between magnetic moments of equal magnitude at the liquid $^3$He–solid state substrate interface becomes possible.

The resonance magnetic coupling between liquid $^3$He nuclei and the $^{141}$Pr nuclei in microsized (45 μm) van Vleck paramagnet PrF$_3$ powder has been discovered by authors [1]. Using nanosized PrF$_3$ powder would create a highly-coupled $^3$He - $^{141}$Pr spin system and could show new aspects of effects discovered earlier. From the other hand, the low temperature magnetism of nanosized PrF$_3$ powders could exhibit new features, compare to the magnetism of bulk van Vleck paramagnet PrF$_3$ due to the great impact of huge surface area of a sample. Besides, the influence of quantum confinement in the case of nanoscopic samples also could give some additional effects. Thus, synthesis of nanosized PrF$_3$ powders and study of their low temperature magnetism is very interesting goal.

The method of synthesis of $PrF_3$ nanoparticles described in [2] was tested earlier [3] by authors. Further development of this method is the main goal of present work. It was in the interest to see the influence of chemical precursors and the time of hydrothermal reaction on the size and the shape of synthesized nanoparticles.

**The Synthesis of nanosized $PrF_3$ samples**

The nanosized $PrF_3$ samples 1, 2, 3 and 4 were synthesized by using the methods described in [2, 3]. In a typical synthesis, 3.72 g of praseodymium oxide was dissolved in 240 ml of a 10% nitric acid $HNO_3$ solution to form a transparent light-green solution

$$Pr_2O_3 + HNO_3 \longrightarrow Pr(NO_3)_3 + H_2O,$$

then, after filtering 2.85 g of NaF (F:Pr=3:1,) was added into the above solution under violent stirring. A light green colloidal precipitate of $PrF_3$ appeared immediately.

$$Pr(NO_3)_3 + NaF \longrightarrow PrF_3 + NaNO_3$$

The pH of the suspension was adjusted by ammonia to about 4.0–5.0. Deionized water was filled into the suspension to make the volume up to 450 ml. After stirring for about 20 min, the suspension was finally transferred into a 500 ml round flask (synthesis of sample 1 has been stopped at this stage) and placed in the microwave oven (650 W, 2.45 GHz) for the further hydrothermal reaction. The suspension was heated by microwave irradiation at 70% of the maximum power under refluxing for 20, 40 and 60 min (samples 2, 3 and 4 respectively). The resulting product was collected by centrifugation (14000 - 17000 RPM) and washed several times using deionized water.

Samples 5 and 6 were synthesized, basically, by eliminating first chemical reaction in the procedure described above, because precursor $Pr(NO_3)_3$ has been used instead of $Pr_2O_3$. In a typical synthesis, 9.8 g of praseodymium nitrate $Pr(NO_3)_3$ and 2.51 g of ammonium fluoride $NH_4F$ was dissolved in 18.75 ml of deionized water.

$$Pr(NO_3)_3 + 3NH_4F \longrightarrow PrF_3 + 3NH_4NO_3$$

A light green colloidal precipitate of $PrF_3$ appeared immediately. Deionized water was filled into the suspension to make the volume up to 450 ml. After stirring for about 20 min, the suspension was finally transferred into a 500 ml round flask (synthesis of sample 5 has been stopped at this stage) and placed in the microwave oven (650 W, 2.45 GHz) for the further hydrothermal reaction. The suspension was heated by microwave irradiation at 70% of the maximum power under refluxing for 20 min (sample 6 has been obtained). The resulting product was collected by centrifugation (14000 - 17000 RPM) and washed several times using deionized water.

**X-ray diffraction (XRD) of synthesized PrF$_3$ samples**

The crystal structure of the samples has been characterized by X-ray diffraction (XRD) (Fig. 1). All of the diffraction peaks can be readily indexed from the standard powder diffraction data of the hexagonal phase PrF$_3$. Figure 1 also confirms the hexagonal phase of the of PrF$_3$ particles crystal structure. As shown in Fig. 1, the narrow and sharp peaks indicate high crystallinity of the samples. Basically, there are no differences in the XRD results between all 6 samples, so chosen different synthesis methods did not affect the crystal structure of nanoparticles, detectable by XRD method.

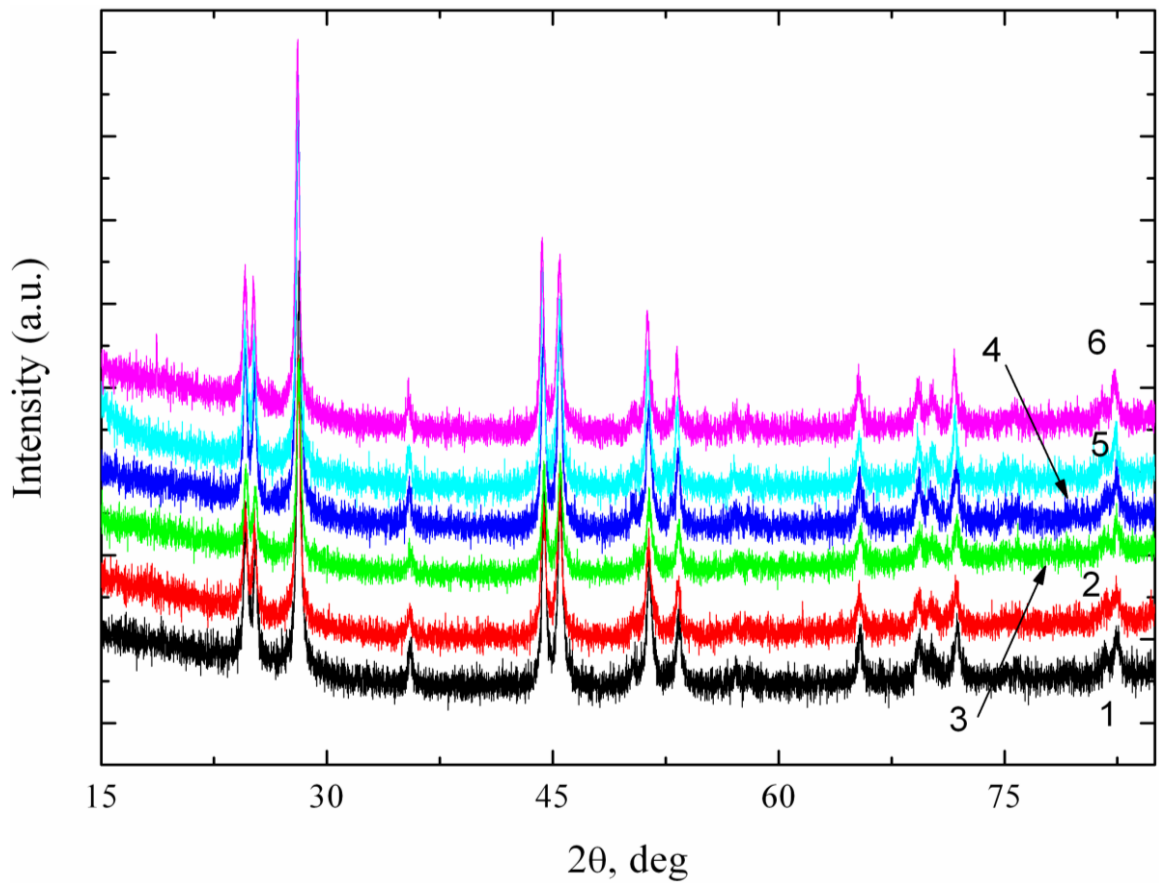

Fig. 1. XRD results of PrF3 samples.

**High-resolution transmission electron microscopy of synthesized PrF$_3$ samples**

High-resolution transmission electron microscopy (HRTEM) images for 6 synthesized PrF$_3$ samples were obtained by using JEM—4000ex with resolution—0.16 nm using an accelerating voltage of 400 kV (Fig. 2 - Fig.7). It is clear, that precursors of chemical reaction have great influence on the shape of synthesized nanoparticles (see difference between shape of samples 1 and 5, as well as between 2 and 6).

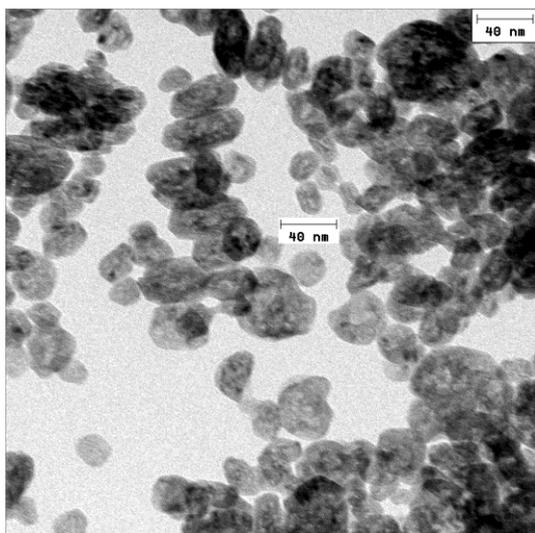 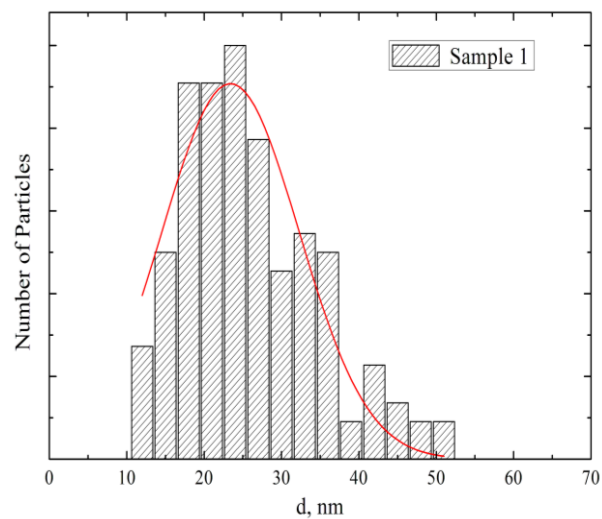

Fig. 2. HRTEM image of PrF$_3$ nanoparticles (sample 1) and size distribution (peak width 18 nm, characteristic size 27 nm, red line is a gauss-line fitting of the distribution)

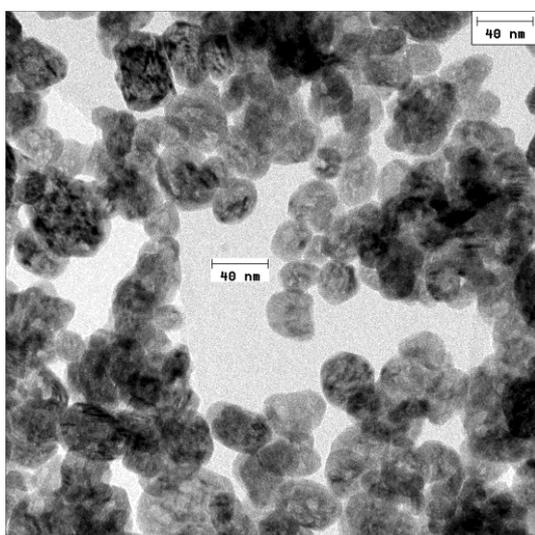 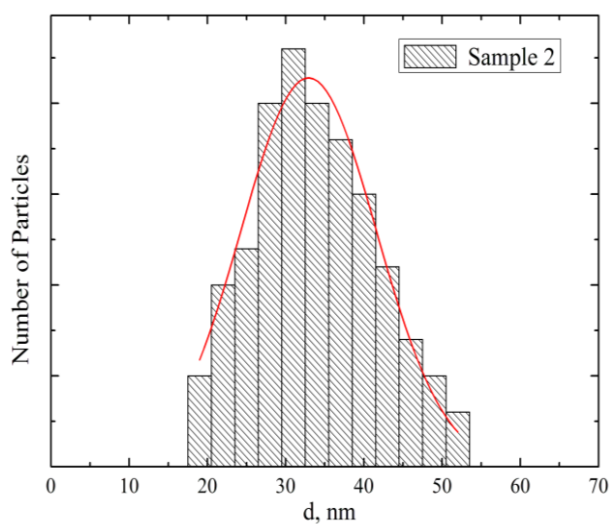

Fig. 3. HRTEM image of PrF$_3$ nanoparticles (sample 2) and size distribution (peak width 17 nm, characteristic size 33 nm, red line is a gauss-line fitting of the distribution)

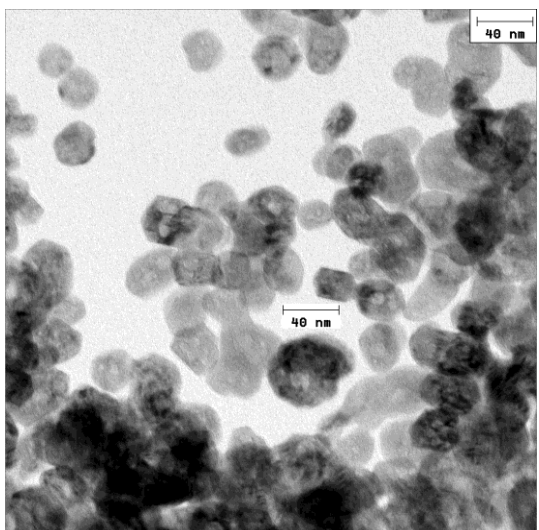 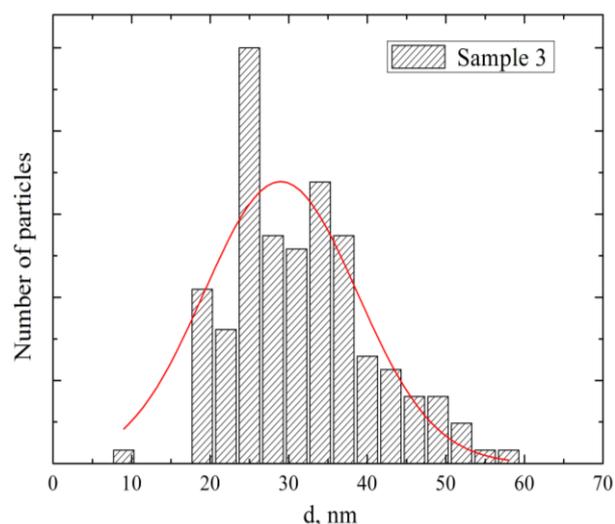

Fig. 4. HRTEM image of PrF$_3$ nanoparticles (sample 3) and size distribution (peak width 19 nm, characteristic size 29 nm, red line is a gauss-line fitting of the distribution)

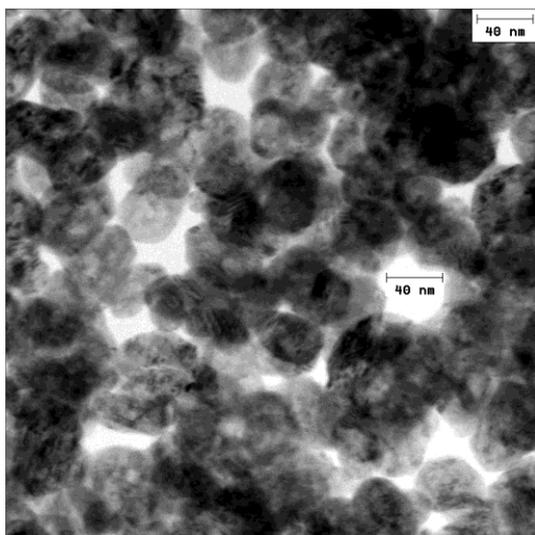
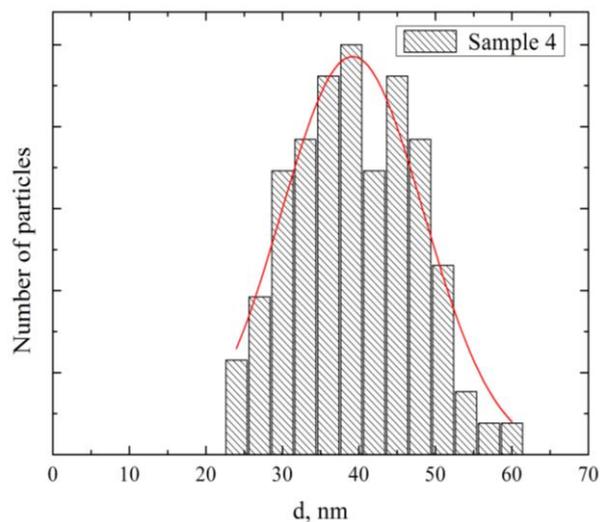

Fig. 5. HRTEM image of PrF3 nanoparticles (sample 4) and size distribution (peak width 19 nm, characteristic size 39 nm, red line is a gauss-line fitting of the distribution)

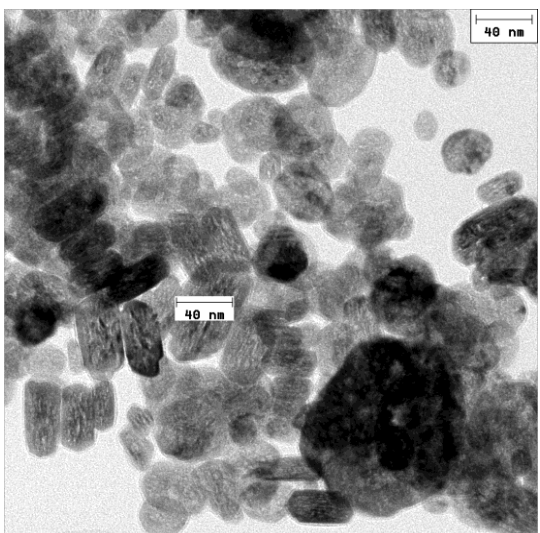
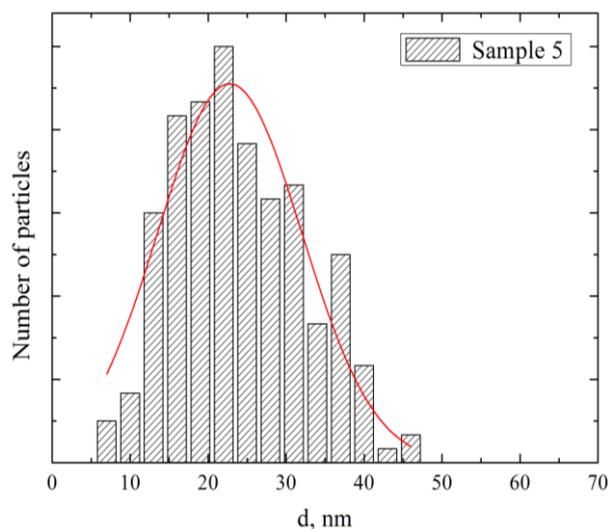

Fig. 6. HRTEM image of PrF3 nanoparticles (sample 5) and size distribution (peak width 18 nm, characteristic size 23 nm, red line is a gauss-line fitting of the distribution)

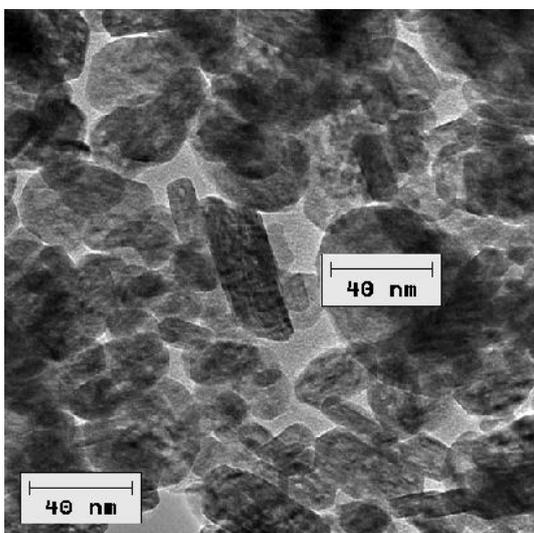
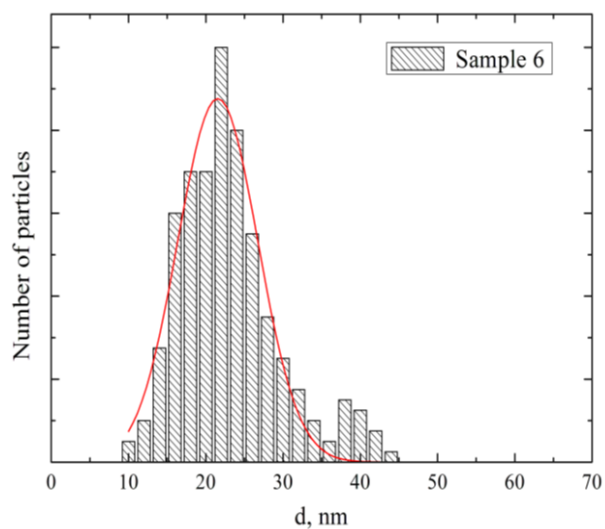

Fig. 7. HRTEM image of PrF3 nanoparticles (sample 6) and size distribution (peak width 10 nm, characteristic size 22 nm, red line is a gauss-line fitting of the distribution)

The dependence of characteristic size for all 6 synthesized samples of the hydrothermal reaction time is presented on the Fig.8 and for samples 1 – 4 it can be characterized by grew rate 10 nm / hour roughly. For samples 5 and 6 there are no dependence, virtually.

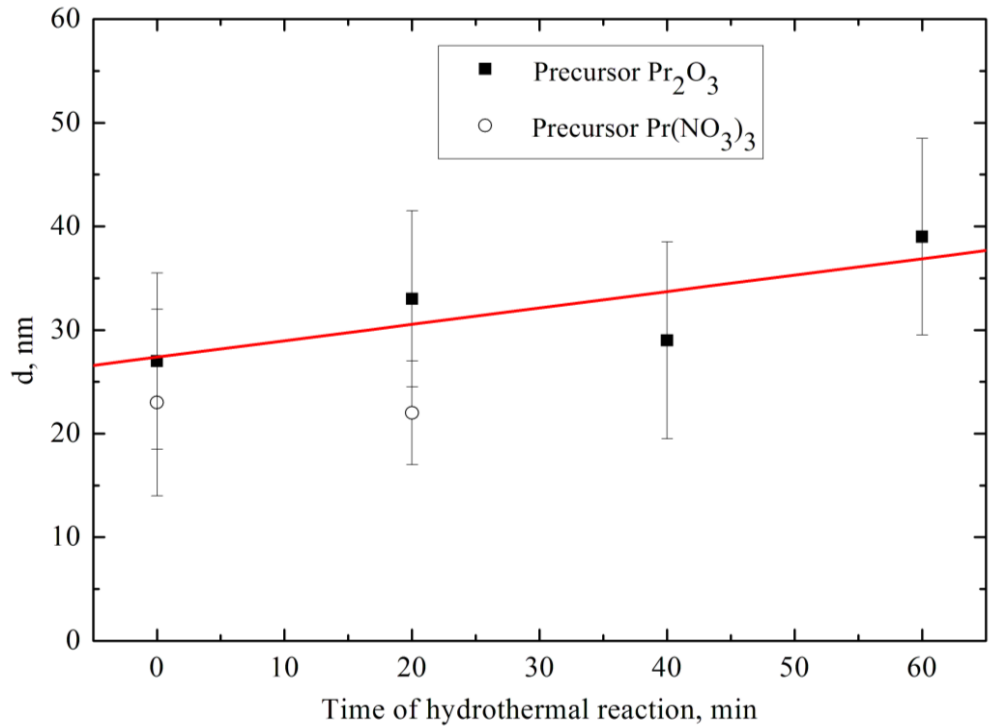

Fig. 8. The dependence of characteristic size of the hydrothermal reaction time

**Conclusion**

The six nanosized $PrF_3$ samples were synthesized using two different chemical reactions and different time of hydrothermal reaction. The X-ray and HRTEM experiments showed high crystallinity of synthesized samples. For all samples the particles size distribution was obtained. It was shown, that precursors of chemical reaction have great influence on the shape of synthesized nanoparticles. The size of nanoparticles depended on the time of hydrothermal reaction as much as roughly 10 nm per hour at our microwave power level.

This work is partially supported by the Ministry of Education and Science of the Russian Federation (FTP "Scientific and scientific - pedagogical personnel of the innovative Russia" GK-P900).

**Acknowlegments**

The authors acknowledge A. Cherkov and A. Gutakovskii for HRTEM images.